\documentstyle[aps,prl,floats,graphicx]{revtex}
\begin{document}

\def\topfraction{1} \def\bottomfraction{1} \def\textfraction{0}

\draft
\title{Vacuum energy and cosmological constant: \\ View from
condensed matter}
\author{G.E. Volovik\\
Low Temperature Laboratory, Helsinki
  University of Technology\\
 Box 2200, FIN-02015 HUT, Finland\\
and\\
L.D. Landau Institute for Theoretical Physics, 117334 Moscow,
  Russia}
\maketitle
\begin{abstract}
The condensed matter examples, in which the effective gravity appears in the
low-energy corner as one of the collective modes of quantum vacuum, provide a
possible answer to the question, why the vacuum energy is so small.
This answer comes from the fundamental ``trans-Planckian'' physics of quantum
liquids. In the effective theory of the low energy degrees of freedom the
vacuum
energy density is proportional to the fourth power of the corresponding
``Planck'' energy appropriate for this effective theory. However, from the
exact
``Theory of Everything'' of the quantum liquid it follows that its vacuum
energy
density is exactly zero without fine tuning, if: there are no external forces
acting on the liquid; there are no quasiparticles which serve as matter; no
space-time curvature; and no boundaries which give rise to the Casimir
effect. Each
of these four factors perturbs the vacuum state and induces a nonzero value of
the vacuum energy density, which is on the order of the energy density of the
perturbation. This is the reason, why one must expect that in each epoch the
vacuum energy density is on the order of the matter density of the
Universe, or/and
of its curvature,  or/and of the energy density of the smooth component -- the
quintessence.
\end{abstract}

\vskip5mm
Talk presented at ULTI Symposium ``ULTRA LOW ENERGY PHYSICS: METHODS AND
PHENOMENOLOGY'', January 10 - 14, 2001, Finland

\tableofcontents

\section{Introduction. The Theory of Everything in quantum liquids.}

The Theory of Everything for quantum liquids and solids -- ``a
set of equations capable of describing all phenomena that have been observed''
\cite{LaughlinPines} in these quantum systems -- is extremely simple. On the
``fundamental'' level appropriate for quantum liquids and solids, i.e. for all
practical purposes, the $^4$He or $^3$He atoms of these quantum systems can be
considered as structureless: the $^4$He atoms are the structureless bosons
and the
$^3$He atoms are the structureless fermions with spin $1/2$. The Theory of
Everything for a collection of a macroscopic number of interacting $^4$He
or $^3$He
atoms is contained in the many-body Hamiltonian written in the second quantized
form:
\begin{equation}
{\cal H}-\mu {\cal N}=\int d{\bf x}\psi^\dagger({\bf
x})\left[-{\nabla^2\over 2m}
-\mu
\right]\psi({\bf x}) +\int d{\bf x}d{\bf y}V({\bf x}-{\bf
y})\psi^\dagger({\bf x})
\psi^\dagger({\bf y})\psi({\bf y})\psi({\bf x})
\label{TheoryOfEverything}
\end{equation}
where $m$ is the bare mass of the atom; $V({\bf x}-{\bf y})$ is the bare
interaction between the atoms;
$\mu$ is the chemical potential -- the Lagrange multiplier which is introduced
to take into account the conservation of the number of atoms: ${\cal
N}=\int d{\bf
x}~
\psi^\dagger({\bf x})\psi({\bf x})$. In $^4$He, the bosonic quantum field
$\psi({\bf x})$ is the annihilation operator of the $^4$He atoms. In $^3$He,
$\psi({\bf x})$ is the fermionic field and the spin indices must be added.

The Hamiltonian (\ref{TheoryOfEverything}) has very restricted number of
symmetries: It is invariant under translations and rotations in 3D space;
there is a global $U(1)$ group originating from the conservation of the
number of
atoms: ${\cal H}$ is invariant under gauge rotation $\psi({\bf x})\rightarrow
e^{i\alpha}\psi({\bf x})$ with constant $\alpha$; in $^3$He in addition, if the
relatively weak spin-orbit coupling is neglected, ${\cal H}$ is also invariant
under separate rotations of spins. At low temperature the phase transition
to the
superfluid or to the quantum crystal state occurs where some of these
symmetries
are broken spontaneously. For example, in the $^3$He-A state  all of these
symmetries, except for the translational symmetry, are broken.

\begin{figure}[!!!t]
\centerline{\includegraphics[width=\linewidth]{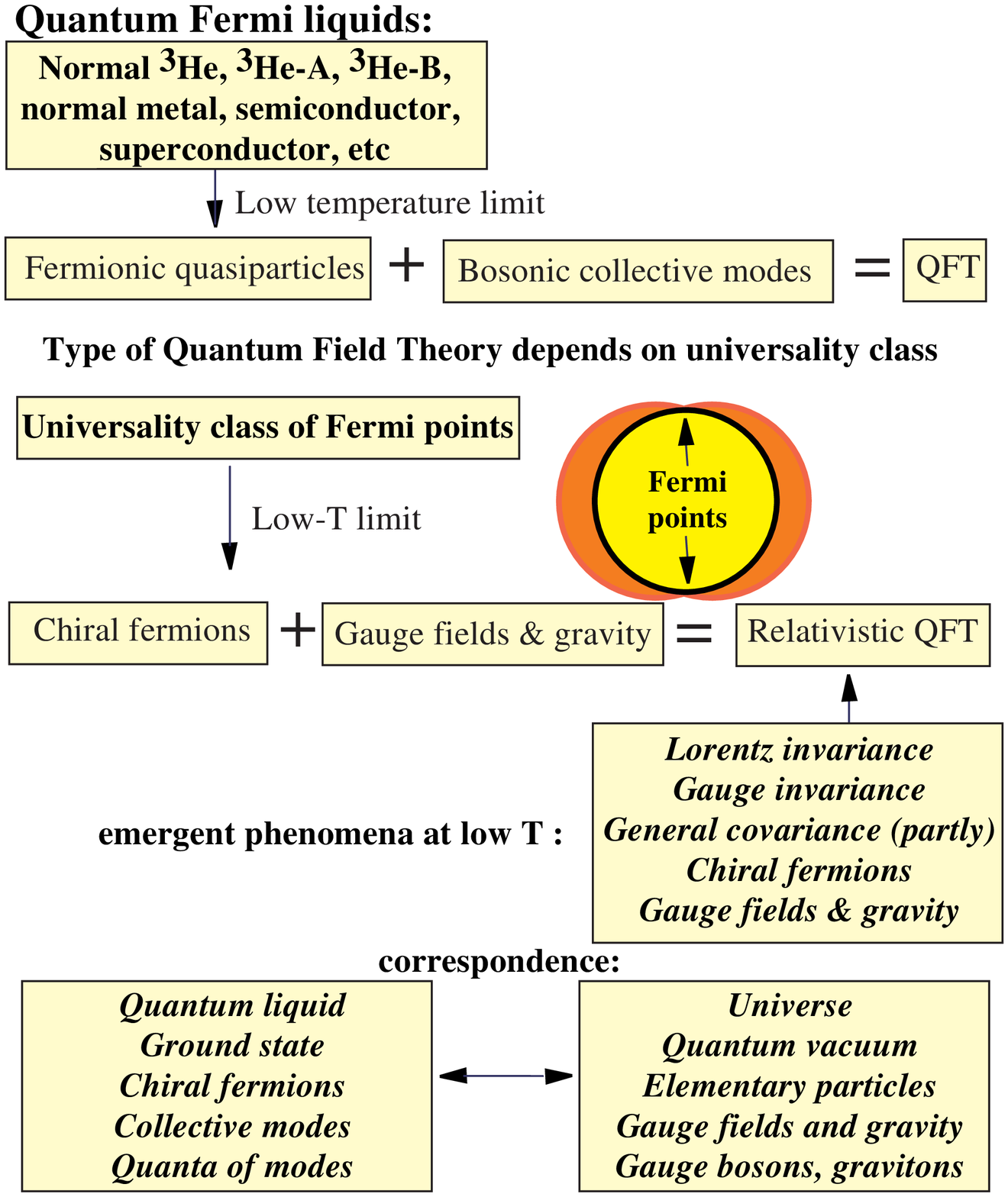}}
\bigskip
\caption[QuantumLiquids]
    {Quantum field theories in quantum liquids. If quantum liquid has Fermi
points, the relativistic QFT emerges at low $T$.}
\label{fig:QuantumLiquids}
\end{figure}

However, when the temperature and energy decrease further the symmetry becomes
gradually enhanced in agreement with the anti-grand-unification scenario
\cite{FrogNielBook,Chadha}. At low energy the quantum liquid or solid is well
described in terms of a dilute system of quasiparticles. These are bosons
(phonons)
in $^4$He and fermions and bosons in
$^3$He, which move in the background of the effective gauge and/or gravity
fields
simulated by the dynamics of the collective modes
(Fig.~\ref{fig:QuantumLiquids}).
In particular, phonons propagating in the inhomogeneous liquid are described
by the effective Lagrangian
\begin{equation}
L_{\rm effective}=\sqrt{-g}g^{\mu\nu}\partial_\mu\alpha \partial_\nu\alpha~,
\label{LagrangianSoundWaves}
\end{equation}
where $g^{\mu\nu}$ is the effective acoustic metric provided by inhomogeneity
and flow of the liquid \cite{Unruh,Stone}.

These quasiparticles serve as the elementary particles of the low-energy
effective
quantum field theory. They represent the analogue
of matter. The type of the effective quantum field theory -- the theory of
interacting fermionic and bosonic quantum fields -- depends on the universality
class of the fermionic condensed matter (see review \cite{PhysRepRev}). In
superfluid $^3$He-A, which belongs to the same universality class as the
Standard
Model, the effective quantum field theory contains chiral ``relativistic''
fermions, while the collective bosonic modes interact with these ``elementary
particles'' as gauge fields and gravity (Fig.~\ref{fig:QuantumLiquids}).
All these
fields emergently arise together with the Lorentz and gauge invariances and
with
elements of the general covariance from the fermionic Theory of Everything in
Eq.(\ref{TheoryOfEverything}).

The emergent phenomena do not depend much on the details of the Theory of
Everything \cite{LaughlinPines}, in our case on the details of the pair
potential
$V({\bf x}-{\bf y})$. Of course, the latter determines the universality
class in
which the system enters at low energy. But once the universality class is
established, the physics remains robust to deformations of the pair
potential. The
details of $V({\bf x}-{\bf y})$ influence only the ``fundamental''
parameters of
the effective theory (``speed of light'', ``Planck'' energy cut-off, etc.)
but not
the general structure of the theory.  The quantum liquids are strongly
correlated
and strongly interacting systems. That is why, though it is possible to
derive the
parameters of the effective theory from first principles in
Eq.(\ref{TheoryOfEverything}), if one has enough computer time and memory,
this is
a rather difficult task. However, in most cases it is appropriate to
consider the
``fundamental'' parameters as phenomenological.

\section{Condensed matter view on cosmological constant problems}

\subsection{Why is the cosmological constant so small?}

The most severe problem in the marriage of gravity and quantum theory is why
the vacuum is not gravitating \cite{WeinbergReview} (see also
Ref.\cite{Polarski}).
The vacuum energy density can be easily estimated: the positive
contribution comes
from the zero-point energy of the bosonic fields and the negative -- from the
occupied negative energy levels in the Dirac sea. Since the largest
contribution
comes from the high momenta, where the energy spectrum of particles is
massless,
$E=cp$, the estimation for the flat space with Minkowski metric
$g_{\mu\nu}={\rm
diag}(-1,c^{-2},c^{-2},c^{-2})$ gives the following energy density of the
vacuum:
\begin{equation}
\rho_\Lambda = {1\over 2V}\sum_{\rm bosons}cp~~ -{1\over V} \sum_{\rm
fermions}cp
\sim
 \pm {1\over c^3} E^4_{\rm Planck} =\pm \sqrt{-g}~E^4_{\rm Planck}~,
\label{VacuumEnergyPlanck}
\end{equation}
where $V$ is the volume of the system. If there is no symmetry between the
fermions
and bosons (supersymmetry) the cut-off is provided by the Planck energy scale
$E_{\rm Planck}\sim 10^{19}$ GeV, with the sign of the vacuum energy being
determined by the fermionic and bosonic content of the quantum field theory. In
case of supersymmetry, the cut-off is somewhat less, being determined by the
scale at which supersymmetry is violated.

If the vacuum energy in Eq.~(\ref{VacuumEnergyPlanck}) is gravitating, this
is in
severe contradiction with the experimental observations, which show that
$\rho_\Lambda$ is less than or on the order of $10^{-120} E^4_{\rm Planck}/c^3$
\cite{Supernovae}. If the vacuum energy is not gravitating, this is in
contradiction with the general principle of equivalence, according to which the
inertial and gravitating masses must coincide.
What can be said about this issue in quantum liquids, where something
similar to
gravity arises in the low energy corner?

Let us first look at the calculation of the vacuum energy. The
advantage of the quantum liquid is that we know both the effective theory there
and the fundamental Theory of Everything in Eq.~(\ref{TheoryOfEverything}).
That
is why we can compare the two approaches. Let us consider for simplicity
superfluid $^4$He.  The effective theory there contains phonons as elementary
bosonic quasiparticles and no fermions. That is why the analogue of
Eq.~(\ref{VacuumEnergyPlanck}) for the vacuum energy is
\begin{equation}
\rho_\Lambda = {1\over 2V}\sum_{\rm phonons}cp \sim
{1\over c^3} E^4_{\rm Debye}=\sqrt{-g}~E^4_{\rm Debye}~,
\label{VacuumEnergyEffective}
\end{equation}
where $c$ is the speed of sound; the ``Planck'' cut-off is now determined
by the
Debye temperature $E_{\rm Debye}=\hbar c/a$  with $a$ the interatomic distance,
which plays the role of the Planck length; $g$ is the determinant
of the acoustic metric in Eq.~(\ref{LagrangianSoundWaves}). If one ignores an
overall conformal factor the space-time interval in acoustic metric provided by
the liquid at rest is $ds^2=dt^2-c^{-2}d{\bf r}^2$, so that $\sqrt{-g}=c^{-3}$.

The disadvantages of such calculations of the vacuum
energy within the effective field theory  are: (i) The result depends on the
cut-off procedure; (ii) The result depends on the choice of the zero from which
the energy is counted: a shift of the zero level leads to a shift in the vacuum
energy. To remove these uncertainties, we must calculate the energy density
of the
ground state exactly, using the Theory of Everything in
Eq.~(\ref{TheoryOfEverything}):
\begin{equation}
\rho_\Lambda={1\over V}<{\rm vac}|{\cal H}-\mu {\cal N}|{\rm vac}> ~.
\label{ExactVacuumEnergy}
\end{equation}
Note that this
energy does not depend on the choice of zero level: the overall shift of
the energy
in
${\cal H}$ is exactly compensated by the shift of the chemical potential $\mu$.

Exact calculation means that not only the low-energy degrees of freedom of the
effective theory (phonons) must be taken into acount, but all degrees of
freedom
of the quantum liquid, i.e. including ``Planckian'' and ``trans-Planckian''
physics. At first glance, this is an extremely difficult task, to calculate an
exact energy of the many-body wave function describing the ground state of the
strongly interacting and strongly correlated system of $^4$He atoms in the real
liquid. Fortunately the result immediately  follows from simple thermodynamic
arguments. If there are no external forces acting on the quantum liquid,
then at
$T=0$ in the limit of infinite liquid volume $V$ one obtains exact
nullification of the energy density:
\begin{equation}
\rho_\Lambda={1\over V}<{\rm vac}|{\cal H}-\mu {\cal N}|{\rm vac}>=0 ~.
\label{ZeroVacuumEnergy}
\end{equation}
The proof is simple. The energy density of the liquid in homogeneous state,
$\epsilon=<{\rm vac}|{\cal H}|{\rm vac}>/V$, is a function of its particle
density $n=N/V$. The pressure $P$ at $T=0$ is
determined in a usual way as
$P=-d(V\epsilon(n))/dV$ where we must take into account that $n=N/V$. Then one
obtains the following equation for the external pressure acting on the liquid:
\begin{equation}
P=-{d\left(V\epsilon\left({N\over V}\right)\right)\over dV}=-\epsilon +
n {d\epsilon\over dn}=-\epsilon+\mu n=-{1\over V}<{\rm vac}|{\cal H}-\mu {\cal
N}|{\rm vac}>=-\rho_\Lambda~.
\label{ProveZeroVacuumEnergy}
\end{equation}
This can be also seen from the hydrodynamic action
for the liquid (see Ref.~\cite{Stone}, where it is shown that the pressure
at $T=0$
is equal to the density of the hydrodynamic action). Note that the relation
connecting vacuum energy and pressure, $\rho_\Lambda=-P$, is exactly the
same as
it comes from the variation of Einstein's cosmological term
$S_\Lambda=-\Lambda
\sqrt{-g}$: i.e.  $T_{\mu\nu \Lambda}=(2/\sqrt{-g})\delta S_\Lambda/\delta
g^{\mu\nu}= \Lambda g_{\mu\nu}$.

In the absence of external pressure,
i.e. at $P=0$, Eq.~(\ref{ProveZeroVacuumEnergy}) gives the zero value
for the energy density of the liquid (or solid) in complete equilibrium at
$T=0$.

\begin{figure}[!!!t]
\centerline{\includegraphics[width=\linewidth]{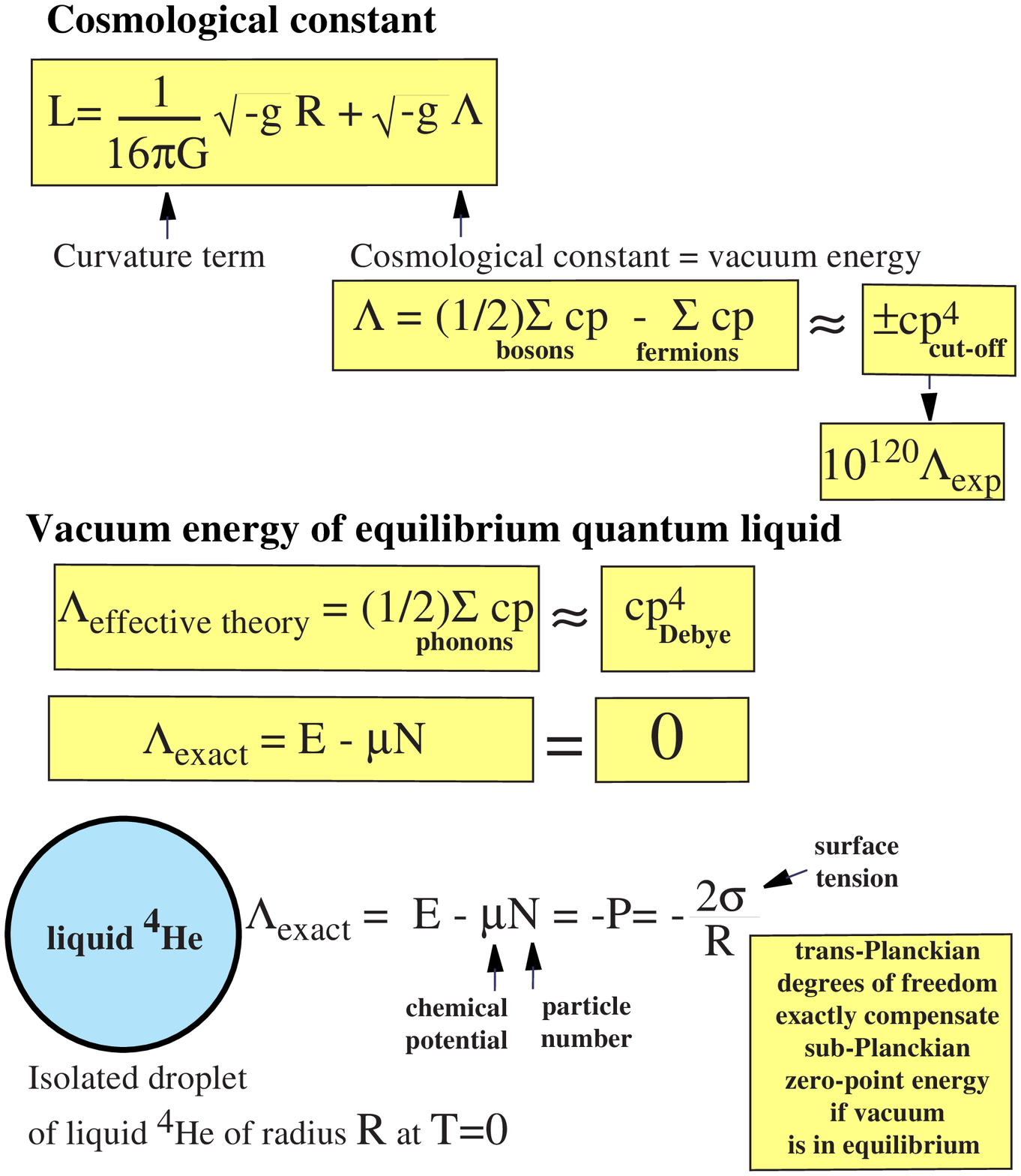}}
\bigskip
\caption[CosmologicalConstant]
    {Vacuum energy and analog of cosmological constant in quantum liquids}
\label{CosmologicalConstant}
\end{figure}

The only condition which we used is that the liquid exists in equilibrium
without external pressure. This condition is fulfilled only for the liquid-like
or solid-like states, for which the chemical potential $\mu$ is negative,
if it is
counted from the energy of the isolated atom. For liquid $^4$He and $^3$He
the chemical potentials are really negative,  $\mu_4 \sim
-7$K and $\mu_3\sim -2.5$K (see review paper
Ref.~\cite{Woo}). This condition cannot be fulfilled for gas-like states for
which
$\mu$ is positive and thus they cannot exist without an external pressure. Thus
the mere assumption that the vacuum of the quantum field theory belongs to the
class of states, which can exist in equilibrium without external forces,
leads to
the nullification of the vacuum energy in equilibrium at $T=0$.

Thus the first lesson from condensed matter is: the standard
contribution to the vacuum energy density from the vacuum
fluctuations in sub-Planckian effective theory is, without any fine tuning,
exactly canceled by the transk-Planckian degrees of freedom, which are not
accessible  within the effective theory.

\subsection{Why is the cosmological constant of order of the present mass of
the Universe?}

We now come to the second problem: Why is the vacuum energy density
presently of
the same order of magnitude as the energy density of matter $\rho_M$, as is
indicated by recent astronomical observations \cite{Supernovae,Polarski}. While
the relation between   $\rho_M$ and $\rho_\Lambda$ seems to depend on the
details
of trans-Planckian physics, the order of magnitude estimation can be readily
obtained. In equilibrium and without matter the vacuum energy is zero.
However, the
perturbations of the vacuum caused by matter and/or by the inhomogeneity of
the metric tensor lead to disbalance. As a result the deviations of the vacuum
energy from zero must be on the of order of the perturbations. Let us
consider how
this happens in condensed matter for different types of perturbations.

\subsubsection{Vacuum energy from finite temperature}

A typical example derived from quantum liquids is the vacuum energy produced by
temperature.  Let us consider for example the superfluid $^4$He in equilibrium
at finite $T$ without external forces. If $T\ll -\mu$ one can neglect the
exponentially small evaporation and consider the liquid as in equilibrium.
Then
the pressure caused by quasiparticles -- phonons -- which play the role of
the hot relativistic matter with equation of state
$P_{M}=(1/3)\rho_M$, must be compensated by the negative vacuum pressure
$P_{\Lambda}=-P_{M}$ to support the zero value of the external pressure,
$P=P_{\Lambda}+P_{M}=0$. In this case one has for the vacuum pressure and
vacuum
energy density
\begin{equation}
\rho_{\Lambda}=-P_{\Lambda}=P_{M}={1\over
3}\rho_{M}=\sqrt{-g}\frac{\pi^2}{30\hbar^3}  T^4~,
\label{VacuumMatterEnergy}
\end{equation}
where $g=-c^{-6}$ is again the determinant of acoustic metric, with $c$
being the
speed of sound. In this example the vacuum energy density
$\rho_{\Lambda}$ is positive and always on the order of the energy density of
matter. This indicates that the cosmological constant is not actually a
constant
but is ajusted to the energy density of matter and/or to the other
perturbations
of the vacuum discussed below.

\subsubsection{Vacuum energy from Casimir effect}

Another example of the induced nonzero vacuum energy density is provided by the
boundaries of the system. Let us consider a finite droplet of $^4$He with
radius $R$. If this droplet is freely suspended then at $T=0$ the vacuum
pressure
$P_{\Lambda}$ must compensate the pressure caused by the surface tension
due to the
curvature of the surface. For a spherical droplet one obtains the negative
vacuum
energy density:
\begin{equation}
\rho_{\Lambda}=-P_{\Lambda}=-{2\sigma\over
R} \sim - {E^3_{\rm Debye}\over\hbar^2c^2 R}\equiv -\sqrt{-g}E^3_{\rm
Planck}{\hbar c\over R}~,
\label{VacuumDropletEnergy}
\end{equation}
where $\sigma$ is the surface tension. This is an analogue of the Casimir
effect,
in which the boundaries of the system produce a nonzero vacuum pressure. The
strong cubic dependence of the vacuum pressure on the ``Planck'' energy $E_{\rm
Planck}\equiv E_{\rm
Debye}$ reflects the trans-Planckian origin of the surface tension $\sigma \sim
E_{\rm Debye}/a^2$: it is the energy (per unit area) related to the
distortion of
atoms in the surface layer of atomic size $a$.  Such term of order $E^3_{\rm
Planck}/R$ in the Casimir energy has been considered in
Ref.\cite{Ravndal}; see also Ref. \cite{Bjorken}, where this vacuum energy is
related to the electroweak vacuum energy.

This form of the Casimir energy -- the surface energy $4\pi R^2\sigma$
normalized
to the volume of the droplet -- can serve as an analogue of the quintessence in
cosmology
\cite{Quintessence,Polarski}. Its equation of state is
$P_{\sigma} =  -(2/3)\rho_{\sigma}$:
\begin{equation}
\rho_{\sigma}={4\pi R^2 \sigma \over {4\over 3} \pi
R^3}={3 \sigma\over R}~~,~~P_{\sigma} =-{2\sigma\over R}= -{2\over 3}
\rho_{\sigma}~.
\label{VacuumDropletEnergy2}
\end{equation}
The equilibrium condition within the droplet can be written as
$P=P_{\Lambda}+P_{\sigma}=0$. In this case the quintessence is related to
the wall -- the boundary of the droplet. In cosmology the quintessence with the
same equation of state,
$<P_{\sigma}> =  -(2/3)<\rho_{\sigma}>$, is represented by a wall wrapped
around
the Universe or by a tangled network of cosmic domain walls \cite{TurnerWhite}.
The surface tension of the cosmic walls can be much smaller than the Planck
scale.

\subsubsection{Vacuum energy induced by texture}

The nonzero vacuum energy density, with a weaker dependence on $E_{\rm
Planck}$,
is induced by the inhomogeneity of the vacuum. Let us discuss the vacuum energy
density induced by texture in a quantum liquid. We consider here
the twist soliton in $^3$He-A. It will be clear later that this texture is
related
to the Riemann curvature in general relativity.
Within the soliton the field of the unit vector $\hat{\bf l}$ changes as
$\hat{\bf l}(z)=\hat{\bf x}\cos\phi(z)+ \hat{\bf y}\sin\phi(z)$. The energy
of the system in the presence of the soliton consists of the vacuum energy
$\rho_{\Lambda}(\phi)$ and gradient energy:
\begin{equation}
\rho= \rho_{\Lambda}(\phi)+\rho_{\rm grad}~,~
\rho_{\Lambda}(\phi)=\rho_{\Lambda}(\phi=0)+{K\over \xi_D^2 }\sin^2\phi~,~
\rho_{\rm grad}=K(\partial_z\phi)^2~,
\label{VacuumEnergySoliton}
\end{equation}
where $\xi_D$ is the so-called dipole length \cite{VollhardtWolfle}.

The solitonic solution of the sine-Gordon equation, $\tan(\phi/2)=e^{z/\xi_D}$,
gives the following spatial dependence of the vacuum and gradient energies:
\begin{equation}
\rho_{\Lambda}(z)-\rho_{\Lambda}(\phi=0)=\rho_{\rm grad}(z) = {K\over \xi_D^2
\cosh^2 (z/\xi_D)}~.
\label{VacuumGradientEnergy1}
\end{equation}
Let us consider for simplicity the 1+1 case. Then the equilibrium state of
the whole quantum liquid with the texture can be discussed in terms of partial
pressures of the vacuum  $P_{\Lambda}=-\rho_{\Lambda}$ and inhomogeneity
$P_{\rm grad}=\rho_{\rm grad}$. The latter equation of state describes the so
called stiff matter in cosmology. In equilibrium the external pressure is
zero and thus the positive pressure of the texture (stiff matter) must be
compensated by the negative pressure of the vacuum:
\begin{equation}
P=P_{\Lambda}(z)+P_{\rm grad}(z)=0~.
\label{VacuumGradientEquilibrium}
\end{equation}
This equilibrium condition produces another relation between the vacuum and the
gradient energy densities
\begin{equation}
\rho_{\Lambda}(z)=-P_{\Lambda}(z)=P_{\rm grad}(z) =\rho_{\rm
grad}(z)~.
\label{VacuumGradientEnergy2}
\end{equation}
Comparison with Eq.~(\ref{VacuumGradientEnergy1}) shows that in equilibrium
\begin{equation}
\rho_{\Lambda}(\phi=0)=0~.
\label{AgainZero}
\end{equation}
As before, the main
vacuum energy density -- the energy density of the bulk liquid far from the
soliton -- is exactly zero if the liquid is in equilibrium. Within the
soliton the
vacuum is perturbed, and the vacuum energy is induced being on the order of the
energy of the perturbation. In this case $\rho_{\Lambda}(z)$ is equal to the
gradient energy density of the texture.

The induced vacuum energy density in Eq.~(\ref{VacuumGradientEnergy1}) is
inversly
proportional to the square of the size of the region where the field is
concentrated: $\rho_{\Lambda}\sim  \sqrt{-g} E^2_{\rm Planck} (\hbar c/
R)^2$ (in
case of soliton $R\sim \xi_D$). Similar behavior for vacuum energy in the
interior
region of the Schwarzschild black hole was discussed in Ref.\cite{Chapline}.

\subsubsection{Vacuum energy due to Riemann curvature}

The vacuum energy $\sim R^{-2}$ has also an analogy in general relativity.
If $R$
is the size of the visible Universe, then since
$E_{\rm Planck}=\sqrt{\hbar c^5/G}$, one obtains  $\rho_{\Lambda}\sim  c^4/GR^2
\sim {\cal  R}/G$, where ${\cal  R}$ is the Riemann curvature.
This analogy with general relativity is supported by the observation that the
gradient energy of a twisted $\hat{\bf l}$-texture is equivalent to the
Einstein
curvature term in the action for the effective gravitational field in $^3$He-A
\cite{PhysRepRev}:
\begin{equation}
{1\over 16 \pi G} \int d^3r  \sqrt{-g}{\cal  R} \equiv K
 \int d^3r
((\hat{\bf l}\cdot(
\nabla\times\hat{\bf l}))^2 ~.
\label{EinsteinActionHe}
\end{equation}
Here ${\cal  R}$ is the Riemann curvature calculated using the effective metric
experienced by fermionic quasiparticles in $^3$He-A,
$ds^2=dt^2 -c_\perp^{-2}(\hat{\bf l} \times d{\bf
r})^2-c_\parallel^{-2}(\hat{\bf
l}\cdot d{\bf r})^2$, with $\hat{\bf l}$ playing the role of the Kasner
axis. That
is why the nonzero vacuum energy density within the soliton
induced by the inhomogeneity of the order parameter is
very similar to that caused by the curvature of space-time.

How the nonzero vacuum energy $\rho_{\Lambda}$ is induced by the spatial
curvature
and matter in general relativity is demonstrated by the solution for the static
closed Universe with positive curvature, obtained by Einstein in his work
where he
first introduced the cosmological term
\cite{EinsteinCosmCon}.  Let us recall this solution.  In the static
state of the Universe two equilibrium conditions must be fulfilled:
\begin{equation}
\rho=\rho_M+\rho_\Lambda+ \rho_{\cal R}=0~,~ P=P_M+P_\Lambda+ P_{\cal
R}=0~.
\label{CurvatureEnergy}
\end{equation}
The first equation in (\ref{CurvatureEnergy}) reflects the
gravitational equilibrium, which requires that the total mass density must be
zero: $\rho=\rho_M+\rho_\Lambda+ \rho_{\cal R}=0$ (actually the
``gravineutrality'' corresponds to the combination  of two equations
in (\ref{CurvatureEnergy}), $\rho+3P=0$, since $\rho+3P$ serves as a source
of the
gravitational field in the Newtonian limit). This gravineutrality is
analogous to the electroneutrality in condensed matter. The second equation in
(\ref{CurvatureEnergy}) is equivalent to the requirement that for the
``isolated''
Universe  the external pressure must be zero:
$P=P_M+P_\Lambda+ P_{\cal R}=0$.  In addition to matter density $\rho_M$ and
vacuum energy density
$\rho_{\Lambda}$, the energy density  $\rho_{\cal R}$ stored in the spatial
curvature is added:
\begin{equation}
 \rho_{\cal R}={{\cal R}\over 16\pi G}=-{3k\over 8\pi GR^2}~,~
P_{\cal R} =-{1\over 3}\rho_{\cal R}~,
\label{CurvatureMass}
\end{equation}
Here $R$ is the cosmic scale factor in the Friedmann-Robertson-Walker metric,
$ds^2=dt^2 - R^2({dr^2\over 1-kr^2} +r^2d\theta^2 +r^2 \sin^2\theta d\phi^2)$;
the parameter
$k=(-1,0,+1)$  for an open, flat, or closed Universe respectively; and we
removed
the factor $\sqrt{-g}$ from the definition of the energy densities.

For the cold Universe with $P_M=0$,  the Eqs.~(\ref{CurvatureEnergy})  give
\begin{equation}
\rho_\Lambda= {1\over 2} \rho_M =-{1\over 3}\rho_{\cal R}= {k\over 8\pi
GR^2} ~,
\label{EinsteinSolution1}
\end{equation}
and for the hot Universe with the equation of state $P_M=(1/3)\rho_M$,
\begin{equation}
\rho_\Lambda=   \rho_M =-{1\over 2}\rho_{\cal R}= {3k\over 16\pi
GR^2}~.
\label{EinsteinSolution2}
\end{equation}
Since the energy of matter is positive, the static Universe is possible only
for positive curvature,  $k=+1$, i.e. for the closed Universe.

This is a unique example of the equilibrium state, in which the vacuum
energy on the order of the energy of matter is obtained within the
effective theory
of general relativity. It is quite probable that the static states of the
Universe
are completely contained within the effective theory and are
determined by Eqs.~(\ref{CurvatureEnergy}), which do not depend on the
details of
the trans-Planckian physics.

However, when the non-stationary Universe is considered, the equation of motion
for $\rho_\Lambda$ must be added, which is beyond the effective theory and must
depend on the Planck physics. The connection to the Planck physics can also
solve
the flatness problem. The Robertson-Walker metric describes the spatially
homogeneous space-time as viewed within general relativity. However, if general
relativity is the effective theory, the invariance under the coordinate
transformations exists only at low energy. For the ``Planck'' observer the
Robertson-Walker metric is viewed as space dependent if $k\neq 0$. That is
why the
condition for the global spatial homogeneity of the Universe both in effective
and fundamental theories is
$k=0$: the Universe must be flat.

\subsection{Why is the vacuum energy unaffected by the phase transition?}

It is commonly believed that the vacuum of the Universe underwent one or
several
broken symmetry phase transitions. Since each of the transtions is
accompanied by
a substantial change in the vacuum energy, it is not clear why the vacuum
energy
is (almost) zero after the last phase transition. In other words, why has
the true
vacuum has zero energy, while the energies of all other false vacua are
enormously big?

The quantum liquid answer to this question also follows from
Eq.~(\ref{ZeroVacuumEnergy}).  For simplicity let us assume that the false
vacuum is separated from the true vacuum by a large energy barrier, and thus it
can exist as a (meta)stable state. Then Eq.~(\ref{ZeroVacuumEnergy}) can
also be applied to the false vacuum, and one obtains the paradoxical
result: in the
absence of external forces the energy density of the false vacuum in
equilibrium must be always the same as the energy density of the true
vacuum, i.e.
it must also be zero. Moreover, this can be applied even to the unstable vacuum
which corresponds to a saddle point of the energy functional, if such a vacuum
state can live long enough.  There is no paradox, however: after the phase
transition to a new state has occured, the chemical potential $\mu$ will
automatically ajust itself to nullify the energy density of the new vacuum.
Thus
in an isolated system (the Universe) the vacuum energy density remains zero
both
above and below the phase transition.

\section{Discussion: Why is vacuum not gravitating?}

We discussed the condensed matter view to the problem, why the vacuum
energy is so
small, and found that the answer comes from the ``fundamental trans-Planckian
physics''. In the effective theory of the low energy degrees of
freedom the vacuum energy density of a quantum liquid is of order $E_{\rm
Planck}^4$ with the corresponding ``Planck'' energy appropriate for this
effective
theory. However, from the exact ``Theory of Everything'' of the quantum liquid,
i.e. from the microscopic physics, it follows that the ``trans-Planckian''
degrees
of freedom exactly cancel the relevant vacuum energy without fine tuning. The
vacuum energy density  is exactly zero, if the following conditions are
fulfilled:
(i) there are no external forces acting on the liquid; (ii) there are no
quasiparticles (matter) in the liquid;  (iii) no curvature or
inhomogeneity; and
(iv) no boundaries which give rise to the Casimir effect. Each of these four
factors perturbs the vacuum state and induces a nonzero value of the vacuum
energy density of order of the energy density of the perturbation. This is the
reason, why one must expect that in each epoch the vacuum energy density is of
order of the matter density of the Universe, and/or of its curvature, and/or of
the energy density of the smooth component -- the quintessence.

The condensed matter analog of gravity provides a natural explanation, why the
effect of cosmological constant is by 120 orders of magnitude smaller than the
result based on naive estimation of the vacuum energy. It also shows how the
effective cosmological constant of the relative order of $10^{-120}$ naturally
arises as the response to different time-independent perturbations. At the
moment
it is not clear what happens in a real time-dependent situation of expanding
Universe, i.e. how the quantum vacuum self-consistently responds to the time
dependent matter field. This requires the knowledge of the dynamics of the
vacuum,
which probably is well beyond the  effective theory of general relativity and
requires some new microscopic Lagrangian (see e.g. \cite{Polarski}).

Let us also mention, that the actual problem for cosmology is not why the
vacuum
energy is zero (or very small), but why the vacuum is not (or almost not)
gravitating.  These two problems are not necessarily related since in the
effective theory the equivalence principle is not the fundamental physical law,
and thus does not necessarily hold when applied to the vacuum energy. That
is why,
one cannot exclude the situation, when the vacuum energy is huge, but it is not
gravitating. The condensed matter provides an example of such situation too.
If one considers the quantum gas instead of the quantum liquid, one finds
that the gas can exists only at positive external pressure, and thus it has the
negative energy density. The translation to the relativistic language gives
a huge
vacuum energy on the order of the Planck energy scale. Nevertheless, even
in this
situation the equilibrium vacuum is not gravitating, and only small
deviations from
equilibrium state are gravitating \cite{PhysRepRev}.

The condensed matter analogy gives us examples of how the effective gravity
appears as an emergent phenomenon in the low energy corner. In these
examples the
gravity is not fundamental: it is one of the low energy collective modes of the
quantum vacuum.  This dynamical mode provides the effective metric (the
acoustic
metric in
$^4$He) for the low-energy quasiparticles which serve as an analogue of matter.
This gravity does not exist on the microscopic (trans-Planckian) level and
appears
only in the low energy limit together with the ``relativistic''
quasiparticles and
the acoustics itself.

The vacuum state of the quantum liquid is the outcome of the microscopic
interactions of the underlying $^4$He or $^3$He atoms. These atoms, which
live in the ``trans-Planckian'' world and form the vacuum state there,  do not
experience the ``gravitational'' attraction experienced by the low-energy
quasiparticles, since the effective gravity simply does not exist at the
micriscopic scale (we neglect here the real gravitational attraction of the
atoms,
which is extremely small in quantum liquids). That is why the vacuum energy
cannot
serve as a source of the effective gravity field: the pure completely
equilibrium vacuum is not gravitating. On the other hand, the long-wave-length
perturbations of the vacuum are within the sphere of influence of the
low-energy
effective theory, such perturbations can be the source of the effective
gravitational field. Deviations of the vacuum from its equilibrium state,
induced
by different sources discussed here, are gravitating.

\section*{ACKNOWLEDGMENTS}

 This work  was supported in part by the Russian
Foundation for Fundamental Research and by European
Science Foundation.

\end{document}